\documentstyle[preprint,prl,aps]{revtex}
\begin{document}
\draft
\title {Physical nature of all the electronic states in the Thue-Morse chain}
\author{Anathnath Ghosh and S. N. Karmakar\cite{kar}}
\address{Saha Institute of Nuclear Physics,\\
1/AF, Bidhannagar, Calcutta-700064, India.}
\date{\today}
\maketitle

\begin{abstract}
	We present an analytical method for finding all the
electronic eigenfunctions and eigenvalues of the aperiodic
Thue-Morse lattice. We prove that this system supports only
extended electronic states which is a very unusual behavior
for this class of systems, and so far as we know, this is the only
example of a quasiperiodic or aperiodic system in which critical
or localized states are totally absent in the spectrum. Interestingly
we observe that the symmetry of the lattice leads to the
existence of degenerate eigenstates and all the eigenvalues
excepting the four global band edges are doubly degenerate. We
show exactly that the Landauer resistivity is zero for all the
degenerate eigenvalues and it scales as $\sim L^2$ ($L=$ system
size) at the global band edges. We also find that the localization
length $\xi$ is always greater than the system size.
\end{abstract}

\narrowtext

	In recent years, intense theoretical activities have
been observed in the literature towards the understanding of
the electronic properties of quasiperiodic and aperiodic systems.
It has been seen that apart from the usual generic features like
the critical wave functions, singular continuous energy spectrum,
scaling properties of the density of states, power-law behavior of
electrical conductivity etc.\ \cite{koh,sok,chao},
these systems can also sustain many
features characteristic of periodic as well as disordered systems.
As for example, in a very recent work Maci\'{a} and Dom\'{i}nguez-Adame 
\cite{mac} have reported a new class of energy eigenstates in the
spectrum of 1D quasiperiodic Fibonacci lattice having properties
characteristic of extended electronic states. But Fibonacci lattice
is the simplest and the most well-studied quasiperiodic model system
since the discovery of icosahedral symmetry in Al-Mn alloy by Shechtman
et al.\ \cite{sch} and the general consensus about its electronic 
structure is that all the eigenstates are critical with Cantor-set
energy spectrum. This kind of diversity of the electronic and
 other properties
are the general characteristics of almost every quasiperiodic or
aperiodic system. A few such instances are the period-doubling 
\cite{chak,bell},
copper-mean \cite{snk}, Rudin-Shapiro \cite{due} and
 Thue-Morse \cite{cha,bov} etc.
 lattices and in these systems critical
as well as extended wave functions can coexist. Moreover, in some of
these cases there is evidence of localized \cite{hua} or chaotic
wave functions \cite{ryu}. There are also notions like the self-similar, 
lattice-like or Bloch-like wave functions regarding the nature of the
eigenstates. The associated spectra can be absolutely continuous,
singular continuous or point singularities, and in many cases the 
spectra seem to be a mixture of these possibilities. All these information
lead to a somewhat confusing situation, and these
controversies can be resolved only if one has clear knowledge of
every eigenvalue and eigenfunction of the system. The analytical answer
to this problem is not very straightforward and in general, every
quasiperiodic or aperiodic system needs separate attention.

	As a first step, in this paper we shall provide an 
instance of an aperiodic system where we have succeeded in overcoming all
the ambiguities in its electronic properties. This is the case of
the well-known Thue-Morse lattice (TM), a deterministic aperiodic
system whose Schr\"{o}dinger equation has been analytically solved to
give a full account of all its eigenvalues and eigenfunctions.
Recent investigations on this system show that it exhibits many contrasting
physical behaviors. The electronic properties of this lattice is
more like those of periodic lattices than the quasiperiodic systems,
while with regard to Fourier  spectrum it has a degree of order
intermediate between disorder and quasiperiodic lattices. The Fourier
spectrum is singular continuous \cite{mer}, whereas the phonon spectrum
is Cantor-set like \cite{axe}. Even if we focus our attention only
on the electronic properties of the TM lattice, there are a number of
papers in the literature showing many apparently contradicting features.
Using the dynamical trace-map \cite{pey} approach Ryu et al. \cite{ryu} have
provided numerical evidence for the existence of critical
and extended states in the TM lattice within the tight-binding model.
From their results it also appears that lattice-like and chaotic
wave functions are possible in the system. La Rocca \cite{rec} 
performed a multifractal analysis of the system and suggested that
the spectrum contains absolutely continuous parts (extended states)
as well as point singularities (localized states). Huang et al. \cite{hua}
have calculated the mean resistance of the TM system and claimed that
it is more localized than the Fibonacci chain. In an interesting
analytical work, Chakrabrarti et al. \cite{cha} have provided the
restrictive conditions on the energy eigenvalues for which the TM
lattice can support extended electronic states. They have shown that
a definite type of short range correlation among the atoms is responsible
for the existence of extended states in the TM lattice, and it is quite
different from the well-known dimer-type correlation \cite{phi}. But
this work does not provide any insight about the entire spectrum 
giving full details of all the eigenstates of the system. Thus it seems 
that a clear understanding of the electronic properties of the TM
lattice is still lacking and the present paper intends to resolve it
on the basis of rigorous mathematical and physical 
considerations.

	The formalism we introduce is based on the
 dynamical trace-map \cite{pey}
of the TM lattice and the commutation properties of the transfer
matrices, and it provides a very simple analytical method for finding
all the eigenvalues and eigenfunctions of the system. This is essentially
a renormalization group approach and it gives a transparent picture
of the nature of the states. The key point of this work is the
association of the eigenvalues obtained by the trace-map technique
with the commutation properties of the transfer matrices and the
underlying physical mechanism is as follows. We can consider the infinite
TM lattice, at any length scale, as built up of two basic clusters and
the transfer matrices for these clusters become identical at certain
eigenvalues of the system. So we can map the
 aperiodic TM lattice effectively
into a periodic lattice,
and it explains why this lattice can support extended electronic
eigenstates. Most surprisingly we observe that the TM lattice
sustains {\em only} extended wavefunctions, a feature quite uncharacteristic
of any quasiperiodic or aperiodic system.  Also, we have
shown exactly that the localization length $\xi$
is always greater than the system size, which again signifies
the delocalized character of all the states. Quite interestingly we have
noticed that the symmetry of the TM lattice manifests itself in a beautiful
way giving rise to the appearance of degenerate states in one-dimension
within the tight-binding formalism. This a very unusual behavior for
any one-dimensional tight-binding system where
degenerate eigenstates are almost forbidden, and the TM lattice can be
considered as an exception. We have seen that apart from the four
eigenvalues corresponding to the global band edges, all other eigenvalues
are doubly degenerate, and thus our work invalidates the findings of
Cheng et al. \cite{mer}. To gain further insight about the nature of
the states, 
we have also studied the transport properties of the electrons. We
have derived exact expression for the Landauer resistivity of the lattice
and shown that all the degenerate states exhibit Bloch-like behavior in
the sense that the Landauer resistivity are zero for these states, while
for the remaining four states the Landauer resistivity scales as $\sim L^2$,
$L$ being the length of the chain.

	The TM sequence $\sigma_n (\bar{\sigma}_n)$ of order $n$ may
be generated by starting with the symbol $A(B)$ and applying the inflation
rule $A\rightarrow AB$ and $B\rightarrow BA$ $n$-times recursively. It
is easy to see that $\bar{\sigma}_n$ is the complement of $\sigma_n$
obtained by interchanging $A$ and $B$ in $\sigma_n$, and they satisfy
the recursion relations $\sigma_{n+1} = \sigma_n \bar{\sigma}_n$ and
$\bar{\sigma}_{n+1} = \bar{\sigma}_n \sigma_n$. Now one can construct a
TM chain by considering the symbols $A$ and $B$ representing two atoms
$A$ and $B$. The electronic properties of such a lattice are generally
modelled by the on-site version of the nearest neighbor tight-binding
Hamiltonian and the corresponding Schr\"{o}dinger equation in Wannier basis
is given by

\begin{equation}
(E-\epsilon_i)\psi_i = \psi_{i+1}+\psi_{i-1} ~,
\label{ham}
\end{equation}
where $\psi_i$ is the amplitude of the wave function at the
$i$th site, $\epsilon_i = \epsilon_A$ or $\epsilon_B$, and $t$
is the hopping integral being set to unity. Now
first we solve the Eqs.\ (\ref{ham}) for the $n$th generation 
chain with the periodic boundary condition, and then consider the
limit $n\rightarrow \infty$ for finding the electronic structure
of the infinite lattice (see Ref.\ \cite{kohm}). Using the
 transfer matrix method
we now show that all the eigenvalues of the TM
chain can be obtained from its trace-map \cite{pey}, which
satisfies the relation
\begin{equation}
\alpha_{n+1} = 
4 \alpha_{n-1}^2 (\alpha_n - 1 ) + 1 ~~~~~~~\mbox{for}~~ n \geq 2, 
\label{trace}
\end{equation}
where $\alpha_n=(\mbox{Tr} M_n)/2$, $M_n$ being the 
transfer matrix of the $n$th generation
TM lattice. If we choose the origin of energy in such a way that
$\epsilon_A =-\epsilon_B = V$, then the initial conditions are
$\alpha_1 = (E^2-V^2-2)/2$ and $\alpha_2 = 2\alpha_1^2 -2V^2 -1$.
We see that $\alpha_n$ is a polynomial in $E$ of degree $2^n$, and
for periodic boundary condition the relation $\alpha_n =1$ determines
all the $2^n$ eigenvalues of the $n$th generation chain consisting
of $N=2^n$ atoms. Let us now recast the trace-map Eq.\ (\ref{trace})
into the following form
\begin{equation}
\alpha_n = 4^{n-2} \alpha_{n-2}^2 \alpha_{n-3}^2 \cdots \alpha_2^2 \alpha_1^2 
		(\alpha_2 -1) + 1.  \label{ntrace}
\end{equation}
This form of the trace-map is quite suggestive and gives a lot of
information about the spectrum. The eigenvalues of the system are
given by the roots of the equations
\begin{equation}
\alpha_1^2=\alpha_2^2=~~\cdots ~~=\alpha_{n-2}^2=0 
~~~\mbox{and} ~~~\alpha_2-1=0.
		\label{roots}
\end{equation}
The equation $\alpha_2-1=0$ actually determines the global band edges
\cite{fan}, because the energies for which $\alpha_2 >1$ are disallowed
as it implies that $\alpha_n >1$. The global band edges are situated
at $E=\pm (\sqrt{V^2+1} \pm 1)$, and Eqs.\ (\ref{roots}) show that
 these are the
 only four non-degenerate
eigenvalues in the spectrum of the TM chain. It is clear from the rest
of the equations in (\ref{roots}) that all the other eigenvalues are doubly
degenerate. The form of Eq.\ (\ref{ntrace}) also suggests that these
degenerate eigenvalues are all distinct. This is due to the fact that
$\alpha_n$ cannot have any of the earlier $\alpha_i$'s (with $i<n$) as
its factor. From the structure of Eq.\ (\ref{ntrace}) it is obvious
that the spectrum of the $n$th generation TM lattice contains all the
eigenvalues of every preceding lower generation chains. In other words,
the eigenvalues of any arbitrary generation TM lattice remain as 
eigenvalues in all the succeeding higher generation chains, and by the method
of induction they also belong to the spectrum of the infinite TM lattice.
In practice, the spectrum of the infinite lattice can be obtained by
making the generation number $n$ sufficiently large, and we get
two global bands (symmetric around $E=0$)
where all the eigenvalues are doubly degenerate 
excepting those corresponding to the four global band edges.

	Knowing all the eigenvalues of the TM lattice, let us now
look into the nature of the eigenstates of the system. From the
symmetry of the TM sequence, we observe that the infinite lattice
can be built up from every pair of unit cells of the form
$(\sigma_n,\bar{\sigma}_n)$, where $\sigma_n(\bar{\sigma}_n)$ is the
$n$th generation TM sequence. If we define the corresponding pair
of transfer matrices as $(T_n,\bar{T}_n)$, then in the basis of 
$(2\times 2)$ identity matrix $I$ and three Pauli matrices $\sigma_x$,
$\sigma_y$ and $\sigma_z$ both $T_n$ and $\bar{T}_n$ can be expressed
into the following form:

\noindent
For odd $n$ ($n \geq 3$), we have
\begin{equation}
\begin{array}{lll}
T_n & = & \alpha_n I +\beta_n \sigma_x +
\gamma_n \sigma_y +\delta_n \sigma_z~, \\
\bar{T}_n & = & \alpha_n I -\beta_n \sigma_x +
\gamma_n \sigma_y +\delta_n \sigma_z~, 
\end{array}
\label{odd}
\end{equation}
where the coefficients satisfy the recursion relations
\begin{eqnarray}
\alpha_n & = &
\alpha_{n-2}^4 +\beta_{n-2}^4 +\gamma_{n-2}^4 +\delta_{n-2}^4 
 +  2 (\gamma_{n-2}^2 + \delta_{n-2}^2) \nonumber \\
 & & \times (3 \alpha_{n-2}^2 +\beta_{n-2}^2) 
 +  2 (\gamma_{n-2}^2 \delta_{n-2}^2 -\alpha_{n-2}^2 \beta_{n-2}^2)~,
 \nonumber \\
\beta_n & = & 
8 \alpha_{n-2} \beta_{n-2} (\gamma_{n-2}^2 + \delta_{n-2}^2)~, \label{rodd}  \\
\gamma_n & = & 
4 (\alpha_{n-2}^2 - \beta_{n-2}^2 +\gamma_{n-2}^2 +\delta_{n-2}^2)
\alpha_{n-2} \gamma_{n-2}~, \nonumber \\
\delta_n & = & 
 4(\alpha_{n-2}^2 - \beta_{n-2}^2 +\gamma_{n-2}^2 + \delta_{n-2}^2) 
\alpha_{n-2} \delta_{n-2}~, \nonumber
\end{eqnarray}
and the iteration starts from the set $\alpha_1=2ab-1$, $\beta_1=b-a$,
$\gamma_1=-i(a+b)$ and $\delta_1=2ab$, where $a=(E-V)/2$ and $b=(E+V)/2$.

\noindent
For even $n$ ($n\geq 4$), we get
\begin{equation}
\begin{array}{lll}
T_n & = & \alpha_n I + \gamma_n \sigma_y + \delta_n \sigma_z~, \\
\bar{T}_n & = & \alpha_n I + \gamma'_n \sigma_y +
				\delta'_n \sigma_z~,
\end{array}
\label{even}
\end{equation}
and the recursion relations of the coefficients are given by
\begin{eqnarray}
\alpha_n & = & 4 \alpha_{n-2}^2	(\gamma_{n-2} \gamma'_{n-2} +
		\delta_{n-2} \delta'_{n-2}) + \nonumber \\
 &  & (\alpha_{n-2}^2+\gamma_{n-2}^2+\delta_{n-2}^2)(\alpha_{n-2}^2 +
		\gamma_{n-2}^{'2}+\delta_{n-2}^{'2})~, \nonumber \\
\gamma_n & = & \alpha_{n-2} (\theta_{n-2}\mu_{n-2} +
		\phi_{n-2}\nu_{n-2})~, \nonumber \\
\delta_n & = & \alpha_{n-2} (\phi_{n-2}\mu_{n-2} -
		\theta_{n-2}\nu_{n-2})~, \label{reven} \\
\gamma'_n & = & \alpha_{n-2} (\theta_{n-2}\mu_{n-2} -
		\phi_{n-2}\nu_{n-2})~, \nonumber \\
\delta'_n & = & \alpha_{n-2} (\phi_{n-2}\mu_{n-2} +
		\theta_{n-2} \nu_{n-2})~, \nonumber 
\end{eqnarray}
where $\theta_{n-2} = 2(\gamma_{n-2}+\gamma'_{n-2})$,
$\phi_{n-2} = 2(\delta_{n-2}+\delta'_{n-2})$,
 $\mu_{n-2} = \alpha_{n-2}^2 + \gamma_{n-2} \gamma'_{n-2} +
\delta_{n-2} \delta'_{n-2}$ and $\nu_{n-2} = \gamma_{n-2} \delta'_{n-2} -
\gamma'_{n-2} \delta_{n-2}$. 
Here the initial set for iteration is $\alpha_2=\alpha_1^2-\beta_1^2+
\gamma_1^2
+\delta_1^2$, $\gamma_2=2(\alpha_1\gamma_1 - i\beta_1\delta_1)$, $\delta_2= 
2(\alpha_1 \delta_1 + i\beta_1 \gamma_1)$, $\gamma'_2= 2(\alpha_1 \gamma_1 +
i\beta_1 \delta_1)$ and $\delta'_2 = 2(\alpha_1 \delta_1 - i\beta_1
 \gamma_1)$.

	First we consider the case of odd $n$. If we set $\beta_n=0$
in Eqs.\ (\ref{odd}), then the matrices $T_n$ and $\bar{T}_n$ become
identical and this is exactly equivalent to the commutation relation
$[T_{n-1},\bar{T}_{n-1}]=0$. The condition $\beta_n=0$ implies that
$\beta_{n-2}=\beta_{n-4}=\cdots =\beta_3=0$ ($\beta_1$ being non-zero)
and thus it also guarantees the equality between $T_m$ and $\bar{T}_m$ for
every $m=3,~5,~7,~\cdots~n$. 
It can be shown from Eqs.\ (\ref{odd}) and (\ref{rodd})
 that the condition
 $\beta_n=0$ actually satisfies
the eigenvalue equation $\alpha_n=1$. In fact, we can reduce 
$\beta_n=0$ as the set of equations $\alpha_1 = \alpha_3 = \alpha_5 =~ \cdots
= ~\alpha_{n-2} = 0$ and $\gamma_1^2 + \delta_1^2 = 0 $, and this is
 effectively
a subset of the eigenvalue equation Eqs.\ (\ref{roots}) obtained
by the trace-map technique. It may be noted that the equation $\gamma_1^2
+ \delta_1^2 = 0 $ is identical to the relation $ \alpha_2 - 1 = 0$,
which gives the global band edges. Now we show that the rest of the equations
in (\ref{roots}) actually correspond to the equality of $T_n$ and $\bar{T}_n$
for even values of $n$. From Eqs.\ (\ref{even}) and (\ref{reven})
 we see that $T_n$ and
$\bar{T}_n$ become identical when $\alpha_{n-2}=0$,
and this condition also satisfies the eigenvalue equation
$\alpha_n=1$. Finally, combining the results for odd and even $n$, we obtain
the equations $\alpha_1 =\alpha_2 =~\cdots ~=\alpha_{n-2} =0$ and 
$\gamma_1^2 + \delta_1^2 =0$ as the conditions for the equality between 
$T_m$ and $\bar{T}_m$, where $m =3,~4,~\cdots ~n$. It should be noticed
that these equality conditions precisely give all the distinct
 eigenvalues of the
system (see Eqs.\ (\ref{roots})),
 though they cannot detect the degeneracy of the levels.
 Nevertheless, it actually proves that at every eigenvalue
of the system, we can always find a pair of unit cells for the TM lattice
which become identical so far as the electronic properties are concerned.
Thus the aperiodic TM lattice effectively maps into various periodic lattices
for energy eigenvalues of the system, and this is the
physical reason why all electronic states of the TM lattice are extended in 
nature. The wave functions 
do not have any Bloch-like periodicity, but they are
extended in the sense that the amplitudes do not decay at infinity.

For the
sake of illustration, in Fig.1 we display $|\psi_n|^2$ as a function of the
site index $n$ for a few selected energies. Here we have used the initial
conditions $\psi_1 =\psi_0 =1$ which is consistent with the periodic
boundary condition \cite{kohm},
 and the figures clearly indicate the extended character 
of the wave functions. It is worth mentioning that we do not find any
chaotic wave functions as reported in Ref.\ \cite{ryu}.
Normally the critical wave functions show power-law divergence \cite{koh},
however, in the case of TM lattice all the states are well-behaved without
any divergence at infinity. In Ref.\ \cite{kazu} K. Iguchi has shown
that the trace map Eq.(\ref{trace}) has no cycle, which implies that the TM
lattice does not support power-law diverging states.
 It is also apparent from the above analysis
that the transition from extended type states to critical or localized
type states cannot be induced in the TM lattice
by changing the Hamiltonian parameters,
a behavior analogous to that of periodic systems.

	Now  we calculate the Landauer resistivity \cite{chao}
 $\rho$ of the TM
lattice. Since we can express the global transfer matrix $M_n$ 
in the form $M^m$ (where $M=$ unit cell transfer matrix and
$m=$ total number of unit cells)
corresponding to every eigenvalue of the system, it can be written in
the close form 
\[
\begin{array}{lll}
M_n & = & I ~~~\mbox{(for $\alpha_n=0$, $n=1,2,3,\cdots$)}~, \\
 & = & 
\left( 
\begin{array}{ll}
1+m\delta_3 & -im\gamma_3 \\
im\gamma_3 & 1-m\delta_3
\end{array}
\right) ~\mbox{(for $\gamma_1^2 + \delta_1^2=0$)}~,
\end{array}
\]
By embedding a finite size TM chain in a periodic array of $A$ type
atoms connected by hopping integrals $t=1$, we obtain the following
expression for the Landauer resistivity \cite{chao},
\[
\begin{array}{lll}
\rho & = & 0 ~~~\mbox{(when $\alpha_n=0$, $n=1,2,3,\cdots$)} \\
 & = & m^2 \delta_3^2 \tan^2(k/2) ~~~\mbox{(when $\delta_1= i\gamma_1$)} \\
 & = & m^2 \delta_3^2 \cot^2(k/2) ~~~\mbox{(when $\delta_1= -i\gamma_1$)}~,
\end{array}
\]
where the wave vector $k$ is given by the relation $2t\cos k =E-\epsilon_A$.
Thus we can further classify the states according to the transport
properties of the electrons in the TM lattice. We see that all the degenerate
states are Bloch-like in the sense that the Landauer resistivity is zero
for these states, while the four states corresponding to the global band
edges are not Bloch-like as in these cases $\rho \sim L^2$, $L$ being the
length of the chain ($L\equiv N=ma$, $a=$ unit cell length). The non-zero
value of $\rho$ at the band edges physically indicates a transition
between exponentially localized (gap) and extended (band) regions \cite{sok}.

	Let us now  calculate the localization length ($\xi$) of the
TM lattice from the expression for the global transfer matrix $M_n$.
 The Lyapunov
exponent \cite{kohm} can be evaluated exactly from the formula $\gamma =
\lim_{N\rightarrow \infty} [N^{-1}\ln ||M_n||]$, where $||~||$ is the modulus
of the matrix. We observe that for all the
states of system size $N$, the localization length 
always has the value $\xi \equiv 1/\gamma = N/\ln 2$,
 and it proves the absence of localization
in the TM lattice \cite{del}. So from the point of view of localization 
of the electrons all states behave identically, whereas in terms of 
their transport properties the band edge states behave quite differently
from the other states.

	To conclude, in this paper
 we have developed an analytical method which gives
full information about the electronic structure of the TM lattice. 
We have shown
that all the electronic states in the aperiodic TM chain are
extended in nature, which
is a very surprising and completely new result. Another interesting 
finding is that it is an
instance of a tight-binding one-dimensional
system which supports degenerate eigenstates.
In this formalism it is also possible to 
determine the Landauer resistivity and localization
length of the system exactly. We hope that this method will  
provide a guideline 
for studying the electronic structure of systems of similar type.

\begin{figure}
\caption{Plot of $|\psi_n|^2$ versus site number $n$. Here $V=0.5$ and
(a), (b) and (c) correspond, respectively, to energies
2.1180339, 1.5 and 1.9955076. All energies are measured in
units of hopping integral $t$.}
\end{figure}

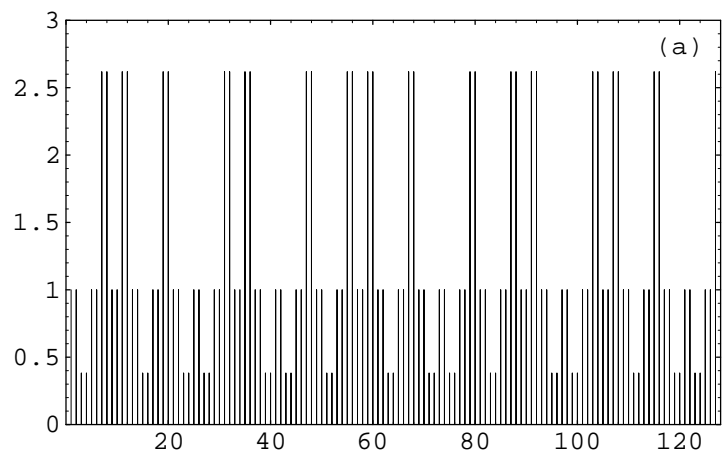
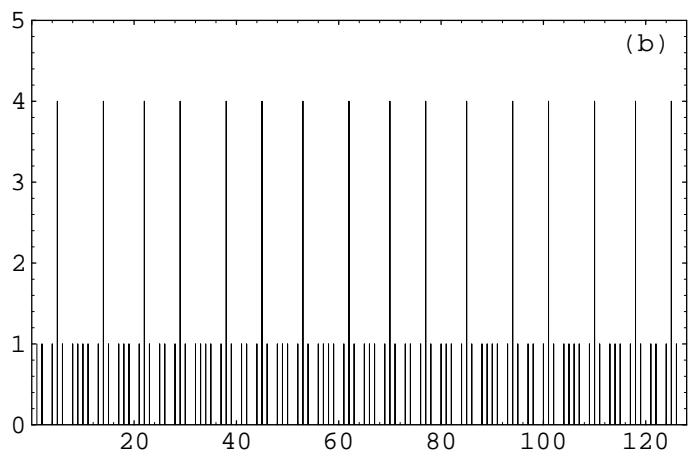
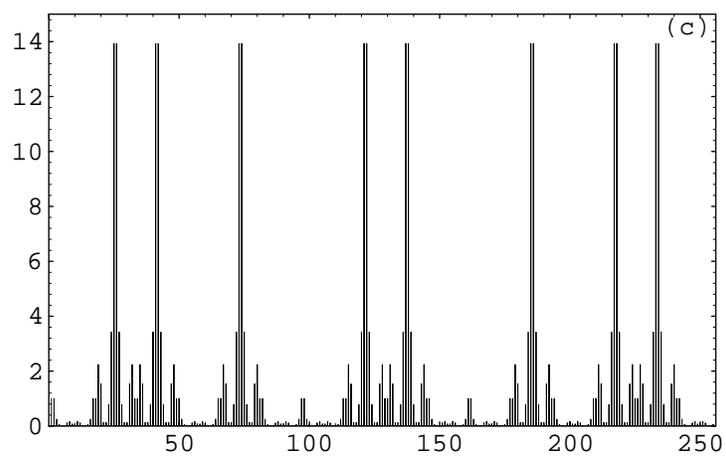
\end{document}